\documentclass[10pt,aps,prl,twocolumn,showpacs,superscriptaddress]{revtex4-1}  
\usepackage{amsmath}
\usepackage{amssymb}
\usepackage{graphicx}
\usepackage{epstopdf}

\begin{document}

\title{Bose-Einstein Condensation of Magnons Pumped by the Bulk Spin Seebeck Effect}

\author{Yaroslav Tserkovnyak}
\author{Scott A. Bender}
\affiliation{Department of Physics and Astronomy, University of California, Los Angeles, California 90095, USA}
\author{Rembert A. Duine}
\author{Benedetta Flebus}
\affiliation{Institute for Theoretical Physics and Center for Extreme Matter and Emergent Phenomena, Utrecht University, Leuvenlaan 4, 3584 CE Utrecht, The Netherlands}

\begin{abstract}
We propose inducing Bose-Einstein condensation of magnons in a magnetic insulator by a heat flow oriented toward its boundary. At a critical heat flux, the oversaturated thermal gas of magnons accumulated at the boundary precipitates the condensate, which then grows gradually as the thermal bias is dialed up further. The thermal magnons thus pumped by the magnonic bulk (spin) Seebeck effect must generally overcome both the local Gilbert damping associated with the coherent magnetic dynamics as well as the radiative spin-wave losses toward the magnetic bulk, in order to achieve the threshold of condensation. We quantitatively estimate the requisite bias in the case of the ferrimagnetic yttrium iron garnet, discuss different physical regimes of condensation, and contrast it with the competing (so-called Doppler-shift) bulk instability.
\end{abstract}



\maketitle

\textit{Introduction}.|The rapidly developing thermoelectric transport capabilities to probe nonconducting materials are instigating a shift in the field of spintronics toward insulating magnets \cite{uchidaNATM10,uchidaAPL10,nakayamaPRL13}. While allowing for seamless spin injection and detection at their boundaries \cite{jiaEPL11,takeiPRB14,chatterjeePRB15}, insulating magnets (including ferromagnets, antiferromagnets, and spin liquids) may offer also efficient spin propagation owing to the lack of the electronic channels for dissipation of angular momentum. Recent measurements of spin signals mediated by thick layers of antiferromagnetic nickel oxide \cite{moriyamaAPL15} and, especially, long diffusion lengths of magnons in ferrimagnetic yttrium iron garnet (YIG) \cite{gilesCM15,cornelissenNATP15,goennenweinCM15}, even at room temperature, bear this view out.

The bosonic nature of magnons, furthermore, naturally lends itself to condensation instabilities when driven by large biases into a nonlinear response \cite{demokritovNAT06,benderPRL12,*benderPRB14,benderCM15}. While the electric spin Hall driving of magnetic insulators \cite{tserkovPRB14,hamadehPRL14,*colletCM15} closely mimics the familiar spin-transfer torque instabilities of conducting ferromagnets \cite{ralphJMMM08}, the possibility of inducing magnonic (Bose-Einstein) condensation also by a heat flux \cite{benderPRL12,*benderPRB14} offers new exciting opportunities that are unique to the insulating heterostructures. The key physics here is played out in the framework of the spin Seebeck/Peltier phenomenology \cite{heremansPHYS14}, according to which the heat and spin currents carried by magnons are intricately intertwined \cite{bauerNATM12}. While the problem of the thermoelectrically-driven magnon condensation has been systematically addressed previously in thin-layer heterostructures \cite{benderPRL12,*benderPRB14,benderCM15}, the more basic regime of an interfacial condensation induced by a bulk heat flux remains unexplored. This concerns the standard geometry of the (longitudinal) spin Seebeck effect, which is suitable for complex lateral heterostructures that could ultimately give rise to useful devices \cite{kiriharaNATM12}.

\begin{figure}[t]
\includegraphics[width=0.9\linewidth]{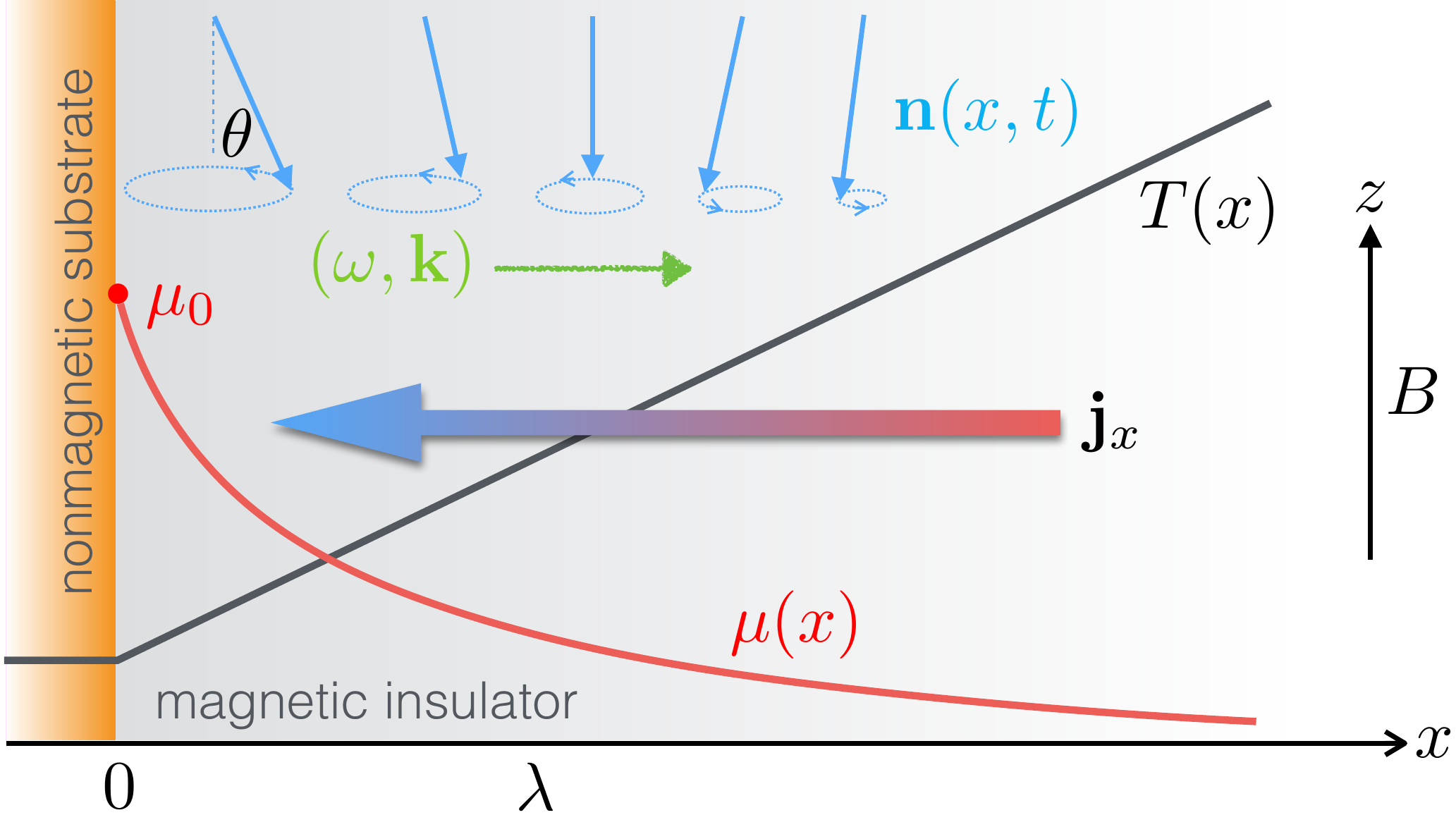}
\caption{A monodomain ferromagnet with uniform equilibrium spin density pointing in the $-z$ direction (in the presence of a magnetic field $B$ pointing up along $z$). A positive thermal gradient, $\partial_xT>0$, induces magnonic flux $\mathbf{j}_x$ towards the interface, where an excess of thermal magnons is accumulating over their spin-diffusion length $\lambda$. When the corresponding nonequilibrium interfacial chemical potential $\mu_0$ reaches a critical value (in excess of the magnon gap), the magnetic order undergoes a Hopf bifurcation toward a steady precessional state, whose Gilbert damping and radiative spin-wave losses are replenished by the thermal-magnon pumping $\propto\mu_0$. The coherent transverse magnetic dynamics decays away from the interface as $n_x-in_y\propto e^{i(kx-\omega t)}$, where ${\rm Im}k>0$.}
\label{sc}
\end{figure}

Applying a large heat flux from a ferromagnet toward its interface with another material (either conducting or insulating), which can carry heat but blocks spin flow,  leads to a nonequilibrium pile up of magnons at the boundary. See Fig.~\ref{sc} for a schematic. When the associated chemical potential of magnons exceeds the lowest-mode frequency of the magnet, the latter gets pumped by the magnonic thermal gas, leading to its condensation at a critical bias. The problem of finding the threshold for this phenomenon as well as considering detrimental and competing effects are the main focus of this Letter. Once experimentally established, such pumped condensates should provide a fertile platform for studying and exploiting spin superfluidity \cite{soninAP10}.

\textit{Two-fluid magnon hydrodynamics}.|The interplay between thermal-magnon transport and coherent order-parameter dynamics is naturally captured within the two-fluid formalism developed in Ref.~\cite{flebusCM15}. Namely, we start with a generic long-wavelength spin Hamiltonian
\begin{equation}
\mathcal{H}=\int d^{3}r \left(- \frac{A}{2s} \hat{\mathbf{s}} \cdot \nabla^{2} \hat{\mathbf{s}} + B\hat{s}_{z} + \frac{K}{2s}\hat{s}^{2}_{z}\right),
\label{H}
\end{equation}
where $\hat{\mathbf{s}}$ is the spin-density operator (in units of $\hbar$), $A$ is the magnetic stiffness, $B$ the external field along the $z$ axis, $K$ the quadratic anisotropy in the same direction (with $K>0$ corresponding to the easy $xy$ plane and $K<0$ easy $z$ axis), and $s$ the saturation spin density. We then perform the Holstein-Primakoff transformation \cite{holsteinPR40} to the bosonic field $\hat{\Phi}\approx(\hat{s}_x-i\hat{s}_y)/\sqrt{2s}$, which is composed of the superfluid order parameter $\Phi\equiv\langle\hat{\Phi}\rangle$ and the quantum-fluctuating piece $\hat{\phi}$: $\hat{\Phi}=\Phi+\hat{\phi}$. These relate to the original spin variables as $\mathbf{s}\equiv\langle\hat{\mathbf{s}}\rangle\approx(\sqrt{2s}{\rm Re}\Phi,-\sqrt{2s}{\rm Im}\Phi,n_c+n_x-s)$, where $\Phi=\sqrt{n_c}e^{-i\varphi}$ and $n_x=\langle\hat{\phi}^\dagger\hat{\phi}\rangle$, with $n_c$ and $n_x$ being respectively the condensed and thermal magnon densities. It is clear that $\varphi$ is the azimuthal angle of the coherent magnetic precession in the $xy$ plane.

Following the Landau-Lifshitz-Gilbert (LLG) phenomenology \cite{landauBOOKv9,*gilbertIEEEM04} of long-wavelength spin-wave dynamics, the following hydrodynamic equations are obtained \cite{flebusCM15}:
\begin{equation}
\dot{n}_x+\boldsymbol{\nabla}\cdot\mathbf{j}_x+\sigma\mu/\lambda^2=2\eta(\omega-\mu/\hbar)n_c\,,
\label{nx}
\end{equation}
for the normal dynamics, where $\mathbf{j}_x =-\sigma\boldsymbol{\nabla}\mu-\varsigma\boldsymbol{\nabla} T$ ($\sigma$ being the magnon conductivity, $\varsigma$ the bulk Seebeck coefficient, and $\mu$ the chemical potential) is the thermal magnon flux and $\lambda$ is the magnon diffusion length, and
\begin{subequations}\begin{align}
\dot{n}_c+\boldsymbol{\nabla}\cdot\mathbf{j}_c+2\alpha\omega n_c&=2\eta(\mu/\hbar-\omega)n_c\,,\\
\hbar(\omega-\Omega)-K\frac{n_c}{s}&=A\left[\left( \boldsymbol{\nabla}\varphi\right)^{2}-\frac{\nabla^2\sqrt{n_c}}{\sqrt{n_c}}\right]\,,
\end{align}\label{nc}\end{subequations}
for the condensate, where $\mathbf{j}_c=-(2A/\hbar)n_c\boldsymbol{\nabla}\varphi$ and $\hbar\Omega=B-K(1-2n_x/s)$ is the magnon gap (where we take for $n_x$ to be the equilibrium cloud density at the ambient temperature $T$ and self-consistently suppose that $\Omega>0$, so that the ferromagnet is in the normal state with $\mathbf{n}=\mathbf{-z}$ in equilibrium \cite{flebusCM15}). Furthermore, $\eta\sim(K/T)^2(T/T_c)^3$ is the dimensionless constant parametrizing the rate of the thermal-cloud|condensate scattering \cite{flebusCM15}, in terms of the Curie temperature $T_c$.

For our present purposes, it will be convenient to recast the condensate dynamics \eqref{nc} in the form of the LLG equation, as discussed in Ref.~\cite{benderCM15}:
\begin{equation}\begin{aligned}
\hbar(1+\alpha\mathbf{n}\times)\dot{\mathbf{n}}&-\left[\hbar\Omega+K(1+\mathbf{n}\cdot\mathbf{z})\right]\mathbf{z}\times\mathbf{n}\\
&=A\mathbf{n}\times\nabla^2\mathbf{n}+\eta\mathbf{n}\times\left(\mu\mathbf{z}\times\mathbf{n}-\hbar\dot{\mathbf{n}}\right)\,,
\end{aligned}\label{llg}\end{equation}
where the second term on the right-hand side is the local thermomagnonic torque parametrized by $\eta$. Rewriting Eq.~\eqref{nx} in the same spirit, we have:
\begin{equation}
\dot{n}_x+\boldsymbol{\nabla}\cdot\mathbf{j}_x+\sigma\mu/\lambda^2=\eta s\,\mathbf{z}\cdot\mathbf{n}\times\left(\dot{\mathbf{n}}-\mu\mathbf{z}\times\mathbf{n}/\hbar\right)\,.
\label{nxl}
\end{equation}

\textit{Spin\;Seebeck-driven instability}.|For the boundary conditions at the interface, $x=0$, we will take the simplest scenario of a hard wall, for which both the thermal and coherent spin currents vanish, leading to
\begin{equation}
\sigma\partial_x\mu+\varsigma\partial_x T=0~~~{\rm and}~~~\partial_x\mathbf{n}=0\,,
\label{bc}\end{equation}
with the latter corresponding to the usual exchange boundary condition for classical ferromagnetic dynamics. Below or near the onset of magnetic instability (condensation in the language of Ref.~\cite{benderPRL12,*benderPRB14}), we can neglect the right-hand side of Eq.~\eqref{nxl}. This produces the spin-diffusion equation, which is solved by
\begin{equation}
\mu(x)=\mu_0e^{-x/\lambda}\,,~~~{\rm where}~~~\mu_0=\lambda\varsigma\partial_x T/\sigma\,,
\label{mu}
\end{equation}
in the steady state (established in response to a uniform thermal gradient $\partial_x T$) and subject to the boundary condition \eqref{bc}. The magnon chemical potential $\mu$ is, naturally, maximized at the interface.

For the remainder of this section, we analyze Eq.~\eqref{llg} subject to the magnonic torque induced by $\mu(x)$ in Eq.~\eqref{mu}. Let us first solve the problem in the limit $\lambda\to\infty$ (relative to other relevant lengthscales, to be identified below) resulting in homogeneous dynamics. Rewriting the corresponding LLG equation \eqref{llg} as
\begin{equation}\begin{aligned}
\hbar(\dot{\mathbf{n}}-\tilde{\Omega}\mathbf{z}\times\mathbf{n})&=\mathbf{n}\times(\eta\mu_0\mathbf{z}\times\mathbf{n}-\tilde{\alpha}\hbar\dot{\mathbf{n}})\\
&\approx(\eta\mu_0-\tilde{\alpha}\hbar\tilde{\Omega})\mathbf{n}\times\mathbf{z}\times\mathbf{n}\,,
\end{aligned}\end{equation}
where $\tilde{\alpha}\equiv\alpha+\eta$ and $\hbar\tilde{\Omega}\equiv\hbar\Omega+K(1+\mathbf{n}\cdot\mathbf{z})$. Here, we assumed $\tilde{\alpha},\eta\ll1$ and thus approximated $\dot{\mathbf{n}}\approx\tilde{\Omega}\mathbf{z}\times\mathbf{n}$ in the Gilbert damping term in going to the second line. It is now easy to see that when the antidamping torque $\propto\eta$ overcomes net damping $\tilde{\alpha}$, the static equilibrium state $\mathbf{n}=\mathbf{-z}$ becomes unstable \cite{ralphJMMM08}. In the case of the easy-axis anisotropy, $K<0$, this leads to magnetic switching toward the stable $\mathbf{n}=\mathbf{z}$ state when $\mu_0>(\tilde{\alpha}/\eta)\hbar\Omega$. In the more interesting easy-plane case, $K>0$ (corresponding to repulsive magnon-magnon interactions), the anisotropy stabilizes magnetic dynamics at a limit cycle (realizing a Hopf bifurcation). The corresponding precession angle $\theta$ is then found to be
\begin{equation}
\theta=2\sin^{-1}\sqrt{\frac{\eta\mu_0-\tilde{\alpha}\hbar\Omega}{2\tilde{\alpha}K}}\,,
\end{equation}
eventually saturating at $\theta\to\pi$ when $\mu_0\geq(\tilde{\alpha}/\eta)(\hbar\Omega+2K)$.

Let us estimate the thermal gradient necessary to reach the critical heat flux for condensation, $\mu_0=(\tilde{\alpha}/\eta)\hbar\Omega$, in the case of yttrium iron garnet. The critical thermal gradient is given by
\begin{equation}
\partial_x T^{(c)}=\frac{\tilde{\alpha}}{\eta}\frac{\sigma}{\lambda\varsigma}\hbar\Omega\,.
\label{dTc}
\end{equation}
Following the magnon-transport theory of Ref.~\cite{flebusCM15} (Supplemental Material), $\sigma/\varsigma\sim1$ \cite{Note1}.
Taking conservatively $\tilde{\alpha}/\eta\sim100$ \cite{benderCM15} and $\lambda\sim10$~$\mu$m \cite{cornelissenNATP15} at room temperature (which is consistent with theoretical estimates based on Ref.~\cite{benderCM15}), we get $\partial_x T^{(c)}\sim1$~K/$\mu$m, for $\Omega/2\pi\sim2$~GHz (corresponding to a kG field). Achieving such thermal gradients should be experimentally feasible \cite{uchidaAPL10,gilesCM15}.

\textit{Condensate outflow}.|More generally, for finite $\lambda$, the condensate is driven near the interface (where $\mu\neq0$) and should eventually decay sufficiently deep into the ferromagnet. This causes spin superflow away from the interface, furnishing radiative spin-wave losses into the bulk, which should suppress condensation and raise the heat-flux threshold. The corresponding instability is described by the LLG equation \eqref{llg}, which we rewrite more compactly as
\begin{equation}
\hbar(\dot{\mathbf{n}}-\tilde{\Omega}\mathbf{z}\times\mathbf{n})=A\mathbf{n}\times\partial_x^2\mathbf{n}+\mathbf{n}\times(\eta\mu\mathbf{z}\times\mathbf{n}-\tilde{\alpha}\hbar\dot{\mathbf{n}})\,,
\label{llgt}
\end{equation}
where both $\tilde{\Omega}$ and $\mu$ become position dependent (both decreasing away from the interface toward $\Omega$ and $0$, respectively, in the bulk). Supposing a smooth onset of instability, we will look for the thermal threshold by setting $\tilde{\Omega}\to\Omega$.

Taking, furthermore, the opposite extreme of $\lambda\to0$ (relative to the absolute value of the condensate wave number, to be checked for internal consistency later), we can integrate Eq.~\eqref{llgt} over a distance $\approx\lambda$ near the interface [noting that $\mathbf{n}\times\partial_x^2\mathbf{n}\equiv\partial_x(\mathbf{n}\times\partial_x\mathbf{n})$] to obtain the boundary condition,
\begin{equation}
\hbar\lambda(\dot{\mathbf{n}}-\Omega\mathbf{z}\times\mathbf{n})\approx A\mathbf{n}\times\partial_x\mathbf{n}+\lambda\mathbf{n}\times\left(\eta\mu_0\mathbf{z}\times\mathbf{n}-\tilde{\alpha}\hbar\dot{\mathbf{n}}\right)\,,
\end{equation}
for the intrinsic bulk dynamics,
\begin{equation}
\hbar(\dot{\mathbf{n}}-\Omega\mathbf{z}\times\mathbf{n})=A\mathbf{n}\times\partial_x^2\mathbf{n}-\tilde{\alpha}\hbar\mathbf{n}\times\dot{\mathbf{n}}\,,
\end{equation}
in the ferromagnet. In order to find the steady-state limit-cycle solution at the onset of the condensation, we linearize these equations with respect to small deviations $\mathbf{m}$ away from equilibrium, $\mathbf{n}\equiv-\mathbf{z}+\mathbf{m}$, and solve for the ansatz $m\equiv m_x-im_y\propto e^{i(kx-\omega t)}$ (requiring that ${\rm Im}k>0$ and $\omega$ is real valued), to obtain
\begin{equation}
\hbar(\omega-\Omega)=Ak^2-i\tilde{\alpha}\hbar\omega\,,
\label{m1}
\end{equation}
subject to the boundary condition
\begin{equation}
i\hbar(\omega-\Omega)=Ak/\lambda-\eta\mu_0+\tilde{\alpha}\hbar\omega\,.
\label{m2}
\end{equation}
The first term on the right-hand side of this equation describes coherent spin outflow into the bulk, the second term magnonic pumping, and the last term Gilbert damping. The critical chemical potential is correspondingly raised as
\begin{equation}
\mu_0=\frac{\tilde{\alpha}\hbar\omega+A\,{\rm Re}k/\lambda}{\eta}\,.
\label{mu0}
\end{equation}
The spin\;Seebeck-induced magnonic pumping $\propto\eta$ thus needs to overcome the condensate outflow $\propto A$ in addition to the Gilbert damping $\propto\tilde{\alpha}$. We proceed to solve Eqs.~\eqref{m1}, \eqref{m2} supposing that ${\rm Im}k\ll\lambda^{-1}$, for internal consistency, and find
\begin{equation}
{\rm Im}k=\left(\frac{\tilde{\alpha}\sqrt{\lambda}}{2\lambda_s^2}\right)^{2/3}\,,\,\,\,{\rm Re}k=\sqrt{\frac{{\rm Im}k}{\lambda}}=\left(\frac{\tilde{\alpha}}{2\lambda\lambda_s^2}\right)^{1/3}\,,
\label{kk}
\end{equation}
where $\lambda_s\equiv\sqrt{A/\hbar\Omega}$ ($\sim10$~nm, using $\Omega/2\pi\sim2$~GHz and typical YIG parameters \cite{bhagatPSS73}). In deriving Eqs.~\eqref{kk}, we have assumed that $\tilde{\alpha}\ll\lambda/\lambda_s$, which should not be an issue in practice. The final internal consistency check is ${\rm Im}k\ll\lambda^{-1}$, which thus boils down to $\tilde{\alpha}(\lambda/\lambda_s)^2\ll1$. For YIG with $\tilde{\alpha}\sim10^{-4}$, this would be borderline when $\lambda/\lambda_s\sim100$ (which should be relevant in practice for a shorter $\lambda$ and/or lower $\Omega$). The frequency according to Eq.~\eqref{m2} is found as $\omega=\Omega(1+{\rm Im}k\,\lambda_s^2/\lambda)\approx\Omega$, so that the instability threshold is finally found according to Eq.~\eqref{mu0} as
\begin{equation}
\partial_x T^{(c)}\approx\frac{\tilde{\alpha}}{\eta}\frac{\sigma}{\lambda\varsigma}\hbar\Omega\left[1+\left(\frac{\lambda^2_s}{\sqrt{2}\tilde{\alpha}\lambda^2}\right)^{2/3}\right]\,,
\label{dTco}
\end{equation}
which is the central result of this Letter.

Note that Eq.~\eqref{dTco} naturally captures also the $\lambda\to\infty$ limit \eqref{dTc} obtained above (thus indicating its general validity for extrapolating between both small and large $\lambda$ regimes), which we now understand as corresponding to $\tilde{\alpha}(\lambda/\lambda_s)^2\gg1$. In the case of YIG at room temperature, we thus expect Eq.~\eqref{dTc} to give a good quantitative estimate for the threshold bias. The details of the magnetic profile beyond the instability threshold can in general be expected to be quite complex, as described by the nonlinear Eq.~\eqref{llgt}, especially if one takes into account the feedback of coherent dynamics on the magnon diffusion according to Eq.~\eqref{nxl}. This nonlinear regime is outside the scope of this work.

\textit{Discussion and outlook}.|At a sufficiently large magnon flux in the bulk of the ferromagnet, the transverse dynamics exhibit also the Doppler-shift instability \cite{bazaliyPRB98,*tserkovPRB06md}, according to the bulk thermomagnonic torque $\propto j_x\partial_x\mathbf{n}$ \cite{kovalevEPL12,*kimPRB15ll}. We find the corresponding threshold to be given by $j_x\sim s\Omega\lambda_s$, which translates into $\partial_xT\sim s\Omega\lambda_s/\varsigma$. Dividing it by the threshold \eqref{dTc}, we get $\partial_xT/\partial_xT^{(c)}\sim (\eta/\tilde{\alpha})(s\lambda_s\lambda/\hbar\sigma)$. Taking \cite{flebusCM15} $\sigma\sim(T/T_c)(s^{2/3}l)/\hbar$, where $l$ is the magnon mean free path, we thus get for this ratio $\sim(\eta/\tilde{\alpha})(T_c/T)(s^{1/3}\lambda_s\lambda/l)$. Performing, once again, an estimate for YIG at room temperature by taking $\eta/\tilde{\alpha}\sim10^{-2}$, $T_c/T\sim2$, $s^{1/3}\sim2$/nm, $l\sim1$~$\mu$m, $\lambda_s\sim10$~nm, and $\lambda\sim10$~$\mu$m, we find that $\partial_xT/\partial_xT^{(c)}\gtrsim1$, so that both instability scenarios are in fact viable and could potentially be competing. This could of course be easily checked as the Doppler-shift instability is independent of the heat-flux direction, while the BEC of magnons discussed here is unipolar, corresponding to the heat flux towards the interface, as sketched in Fig.~\ref{sc}.

It needs also be stressed that the ratio $\eta/\tilde{\alpha}\sim~10^{-2}$ employed in this Letter for our estimates corresponds only to thermal magnons and disregards low-energy magnons that are beyond the Bose-Einstein thermalization description \cite{flebusCM15,benderCM15}. When $\mu_0$ approaches and ultimately exceeds the magnon gap $\hbar\Omega$, the overpopulation of magnons pumped at the bottom of the spectrum could effectively enhance this factor, approaching $\eta/\tilde{\alpha}\to1$ in the extreme case (realizing the limit of the strong condensate-cloud coupling studied in Ref.~\cite{benderPRL12}). This innately nonequilibrium regime, which would yield a lower threshold for magnonic condensation, is, however, beyond our present formalism.

Once established, the interfacial condensate of magnons can be readily detected by monitoring the spin accumulation (utilizing, for example, the magneto-optic Kerr effect) in the adjacent metallic (nonmagnetic) substrate or detecting the associated spin pumping by the inverse spin Hall effect (as in the conventional spin Seebeck geometry \cite{uchidaAPL10}). In the latter case, the theory would have to be complemented with the appropriate treatment of spin leakage into and relaxation in the normal metal \cite{flebusCM15}. The condensate can also be used as a starting point to study and exploit collective ``conveyor-belt" heat and spin flow \cite{flebusCM15} tangential to the interface, which would reflect its superfluid nature.

\acknowledgments
 
The authors thank Joseph P. Heremans and Roberto C. Myers for helpful discussions and experimental motivation for this work. This work is supported by the Army Research Office under Contract No.~911NF-14-1-0016, NSF-funded MRSEC under Grant No.~DMR-1420451, US DOE-BES under Award No.~DE-SC0012190, and in part by the Stichting voor Fundamenteel Onderzoek der Materie (FOM).

\end{document}